# Orientation relationship of FeNiC and FeNiCSi from variant detection in EBSD data

**Mattis Seehaus**[*,1], **Risheng Pei**[1], **Sandra Korte-Kerzel**[1], **Stefanie Sandlöbes-Haut**[1]

[1]Institut für Metallkunde und Materialphysik, RWTH Aachen University, 52074 Aachen, Germany, 2022
[*]Corresponding author, email seehaus@imm.rwth-aachen.de

## Abstract

The determination of orientation relationships in dual or multi-phase materials is very important in the field of interface engineering for the design of materials with tailored properties. In this work, a code is developed for the automated and statistical analysis of the orientation relationship of electron backscatter diffraction data. On the example of Fe-Ni-(Si)-C alloys containing lenticular martensite and retained austenite, the code is applied and it is shown that the orientation relationship (OR) corresponds to the Greninger-Troiano OR and that a statistically reliable investigation of the OR between the retained austenite and the related martensite variants is feasible using the code developed in this study.

# 1 Introduction

In the field of interface engineering, it is important to determine and identify the present orientation relationships (OR) between the different phases in dual or multi-phase materials. This helps to understand and tailor the material properties such as fracture toughness, strength, ductility and the mechanisms of (co-)deformation [1, 2]. An example for materials with distinct ORs are steels containing martensite (bct) and austenite (fcc), whereby the character of the interphase boundary of the constituent phases affects the mechanical properties of the material. In these steels, the OR between neighbouring austenite and martensite grains is mostly investigated by transmission electron microscopy (TEM) due to the usually small grain sizes [3-7]. However, the investigation by means of electron backscatter diffraction (EBSD) has been established in the last decade as a sufficiently performant method to investigate large, representative areas with respect to crystallographic information such as crystal structure (phase), grain orientation or misorientation across grain boundaries in martensitic, low-carbon or duplex steels [8-13].

The EBSD orientation maps, revealing the spatial orientation distribution, are recorded using a scanning electron microscope (SEM) and have the advantage of a higher statistical significance of the discovered ORs compared to the acquisition of individual orientation measurements in the TEM. The orientation maps obtained in this way contain a large amount of crystallographic data, which then leads to a statistically reliable analysis. Prior studies on the determination of ORs between martensite and austenite have revealed some advantages of EBSD over TEM, but have been carried out on individual laths only [14-16]. By reducing the acceleration voltage and / or using transmission Kikuchi diffraction (TKD) on thin (~100 nm) lamellae, the interaction volume between the electron beam and the sample can be reduced, enabling the measurement of small grains similar to TEM [17]. In contrast to the statistically low relevance of the TEM data, the large amount of orientation data obtained by EBSD provides the possibility to determine the differences between experimental pole figures and theoretical pole figures calculated from the respective OR models [18].

A basic prerequisite for such investigations is an OR model. Typically, the OR models define lattice directions and lattice planes. The statistically relevant identification of ORs using austenitic-martensitic steels as an example is shown in this work. Examples for well-established OR models for these steels are the Kurdjumov-Sachs (KS) and Nishiyama-Wassermann (NW) orientation relationships, that were discovered in low-carbon steels on a single austenite grain after the martensitic transformation and in iron-nickel alloys (30 % Ni), respectively [19-21].

These models appear in many studies to classify the OR of martensitic steels [10, 22]. Plate martensite forms at temperatures down to those of liquid nitrogen, unlike lath martensite, resulting in a morphology of accommodated twins exhibiting the Greninger-Troiano (GT) OR [15, 23-27]. However, the KS and NW OR models are considered untenable based on the high indexed martensite habit planes and, moreover, numerous measurements show significant deviations from the ideal model conditions, which could still be referred to as near KS or near NW ORs. [7, 28]. In Table 1 the main ORs proposed or observed are summarized [14, 15, 29].

*Table 1 Observed orientation relationships (ORs) between austenite, γ, and martensite, α', in steels.*

| OR | Plane | Direction |
|---|---|---|
| Bain | $\{100\}_\gamma // \{100\}_\alpha$ | $<100>_\gamma // <110>_\alpha$ |
| NW | $\{111\}_\gamma // \{110\}_\alpha$ | $<112>_\gamma // <110>_\alpha$ |
| KS | $\{111\}_\gamma // \{110\}_\alpha$ | $<110>_\gamma // <111>_\alpha$ |
| GT | $\{111\}_\gamma // \{110\}_\alpha$ | $<123>_\gamma // <133>_\alpha$ |
| Pitsch | $\{100\}_\gamma // \{110\}_\alpha$ | $<110>_\gamma // <111>_\alpha$ |

In the present work, the OR between austenite and martensite is determined using TKD and EBSD in Fe-Ni-C-(Si) alloys, where lenticular martensite and retained austenite co-exist at room temperature. In steels, Si increases the strength and also the amount of retained austenite. Further, Si addition also increases the temperature at which stable C clusters can form, which prevents carbide precipitation like cementite during tempering [30-32]. A code was developed to automatically determine the ORs from the EBSD and TKD data. As a result of the large data sets resulting from these EBSD and TKD mappings, a statistically reliable determination with regard to the martensite/austenite ORs is feasible.

# 2 Experimental procedure

The Fe-24wt%Ni-0.4wt%C and Fe-24wt%Ni-0.4wt%C-2wt%Si alloy samples investigated (hereafter referred to as FeNiC and FeNiCSi, respectively) have been produced by casting, hot rolling at 1050 °C, subsequent cold rolling and recrystallization treatment at 950 °C. At room temperature, these two materials are completely austenitic. The martensitic transformation takes place at temperatures below 0 °C, which facilitates the investigation of the austenitic microstructure at RT before the martensitic transformation. Quenching of only half of the samples in liquid nitrogen preserves a high proportion of austenite, which further enables the determination of the orientation relationship between α' and γ. Samples for microstructure characterisation were cut from the rolled sheet by electrical discharge machining, mechanically ground, polished to 0.25 μm with diamond suspension and finally electrolytically polished with A2 electrolyte at 30 V for 30 s.

SEM and BSE images were taken to investigate the microstructure of the two alloys. Different substructures of the lenticular martensite are clearly visible caused by the presence and absence of silicon as shown in Fig. 1. Characteristic for lenticular martensite is the midrib, which consists of a structure of fine transformation twins from which growth to the lenticular form is initiated.

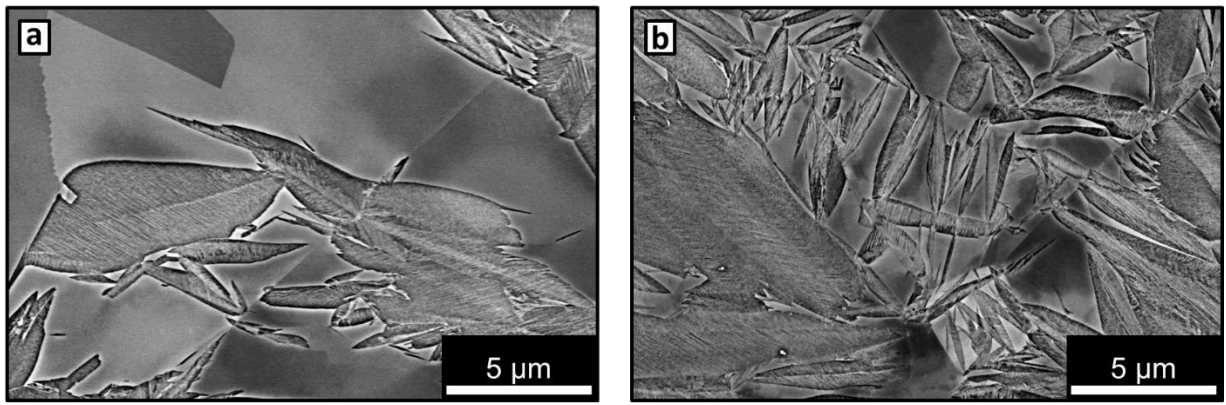

*Fig. 1 Microstructure of the lenticular martensite in a) FeNiC and b) FeNiCSi by means of backscatter electron imaging, respectively.*

EBSD was applied to perform microstructure analysis in FEG-SEM (Helios NanoLab 600i, FEI Co) at 15 kV and 1.4 nA (Hikari, EDAX Inc.). The EBSD maps cover an area of 40 x 30 μm², were recorded with a step size of 150 nm, and were analyzed using the free Matlab® toolbox MTEX 5.7.0 [33].

## 2.1 Code development

For the identification of ORs, a code was developed that uses various methods for analysing two-phase or multi-phase materials in terms of orientation relationships and is accessible as supplementary material and at [34]. In this work, the code is used taking martensite and austenite in Fe-Ni-C(-Si) steels as an example.

In order to determine the orientation relationship between austenite and martensite as well as to compare them, the austenite grains were detected from the EBSD maps of the partially transformed samples and their orientation matrices were rotated according to the standard projection along (001). In conjunction with this orientation transformation, the transformed martensite grains were selected and rotated with the corresponding rotation matrices with reference to the sample symmetry. Several different workflows are enabled and visualised in Fig. 2.

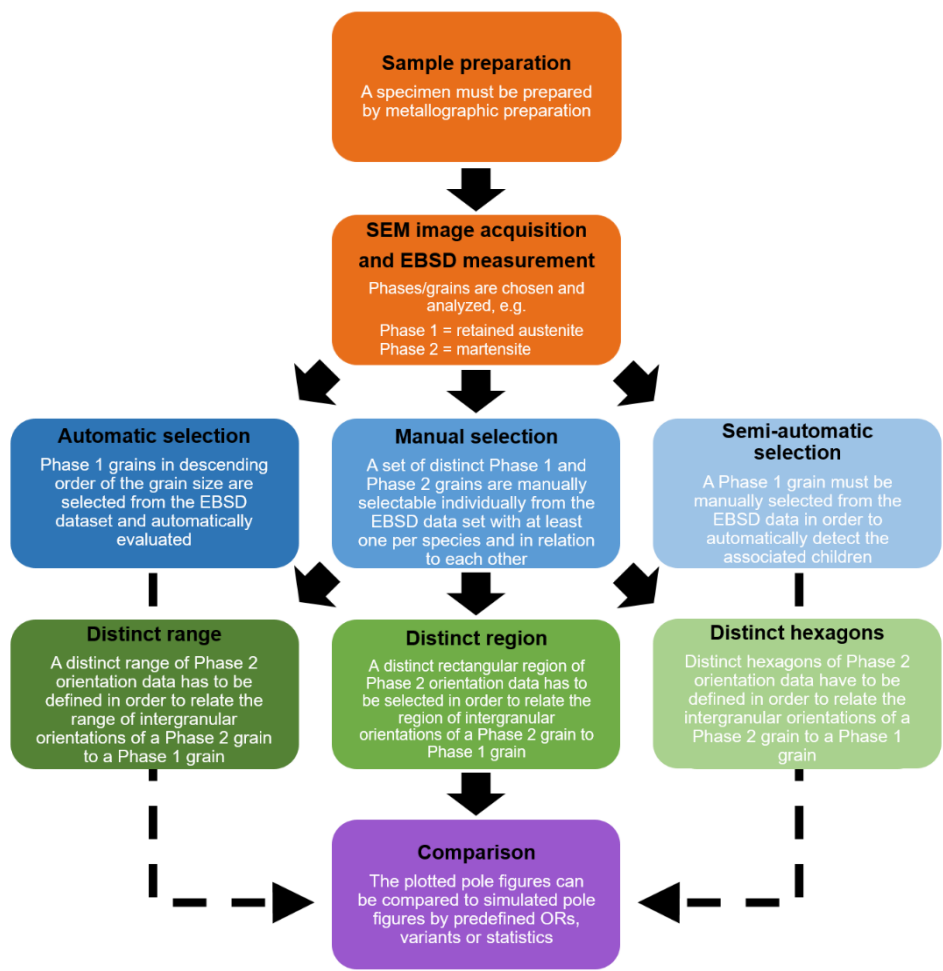

*Fig. 2 Workflow of the OR analysis code. Orange segments show preceding preparations and measurements, blue shows modes 1, 2 and 3 whereas green illustrates the three submodes of mode 2 and purple depicts potential analysis after the code has been executed.*

The developed Matlab code enables the fully automated characterisation of an entire EBSD map and plotting of the result as pole figures. If only specific grains or individual interfaces of an EBSD are of interest, the manual selection of grains is also possible. Furthermore, the different modes can be used to correlate the orientations within a Phase 1 (parent phase) grain or individually selected EBSD data points with the Phase 2 (transformed) grain orientations.

In the automated mode, the EBSD data set is first arranged in descending order of the size of the phase 2 grains and the orientations of a predefined number of the largest grains are stored. This selection is expanded to all austenite grains that exhibit the same orientations within a deviation angle of $\varphi=5°$ to improve statistics. Furthermore, the adjacent Phase 1 grains sharing a grain boundary with the filtered Phase 2 grains are assigned to them and their orientations are stored. Consequently, the Phase 2 grains are rotated to the standard orientation and the corresponding transformation matrices are stored in order to rotate the corresponding Phase 1 grains accordingly. In Fig. 3 an example of an EBSD map of martensite and austenite, the corresponding phase map and the combined EBSD/phase map after completion of the code analysis are depicted.

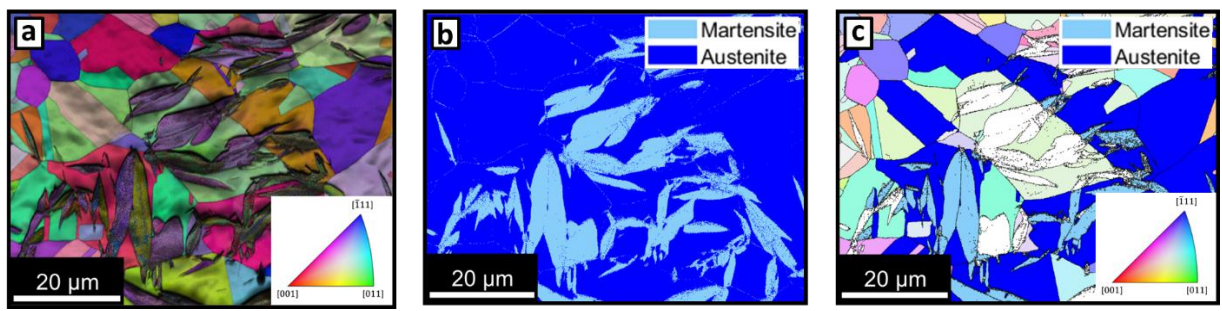

*Fig. 3 a) EBSD map of FeNiC revealing the orientation data b) Phase map showing martensite/austenite grains, respectively and c) combined EBSD/Phase map revealing both the analysed selected austenite grains with the corresponding martensite grains in terms of their OR and the remaining unexamined grains according to their orientation.*

Moreover, it is feasible to select individual Phase 2 grains and either automatically determine the respective fraction of (converted) Phase 1 grains originating from these grains, or to select certain other phase 1 grains and display them correlatively in the pole figures. In Fig. 4a) such a combined EBSD/phase map is depicted with the three individual modes for partial intergranular orientation relationship plots. In Fig. 4b) the intergranular orientations of a single martensite (green box in the extracted region of Fig. 4c)) are shown. Based on the adaptive selection method, the orientations of the individual martensite grains can be compared and plotted as a predefined patterns (Fig. 4d)), as a manually created rectangular area or as individual EBSD data points with the orientations of the parent phase, depending on the specific use case. Using this mode, it is feasible to analyse the OR in specific areas close to interfaces, as the ORs at individual interfaces may vary.

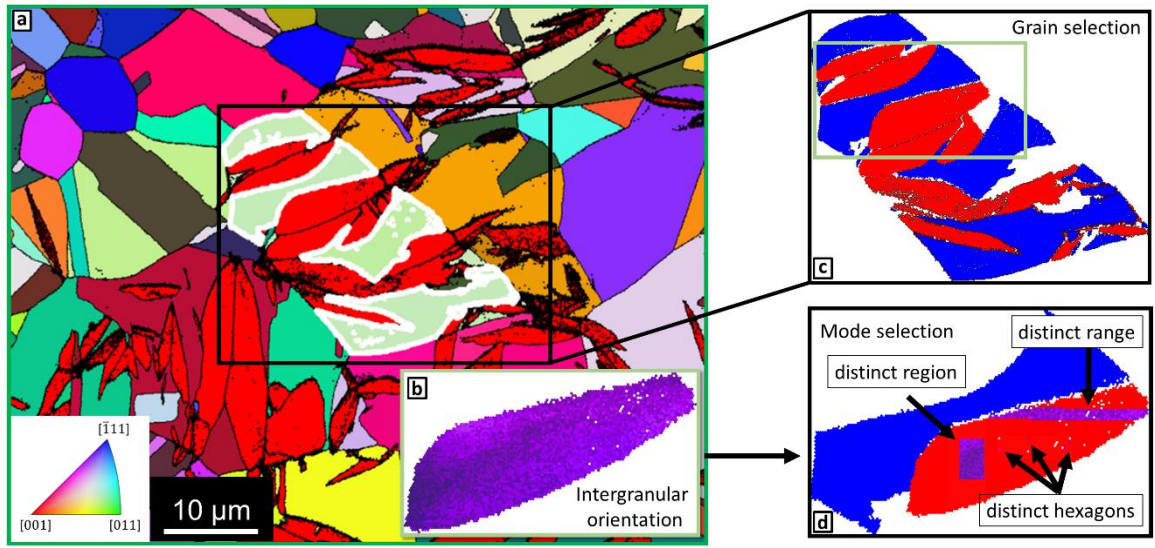

*Fig. 4 a) Combined EBSD and phase map of FeNiC representing the mean austenite orientations as indicated by the inverse pole figure colour code with reference to the norm direction. The martensite is represented in red colour only. b) intergranular orientations of martensite and the corresponding austenite grain in c); d) distinct predefined range, individually selected rectangular area and individually predefined data points of the EBSD data within the martensite and the corresponding austenite grain. The white spots indicate non-indexed points.*

# 3 Results

The procedures available in the code are demonstrated by using FeNiCSi, however, the analysis results of FeNiC obtained using the same procedures will be also presented in the final sections. Fig. 5 shows the measured orientations of martensite and austenite of FeNiCSi and also simulated pole figures of the possible ORs (Table 1), enabling the direct comparison between experimentally measured and simulated orientation relationships, here using the (001), (011) and (111) pole figures. The blue orientations in the pole figures of FeNiCSi correspond to the austenite grains after rotation into the reference position.

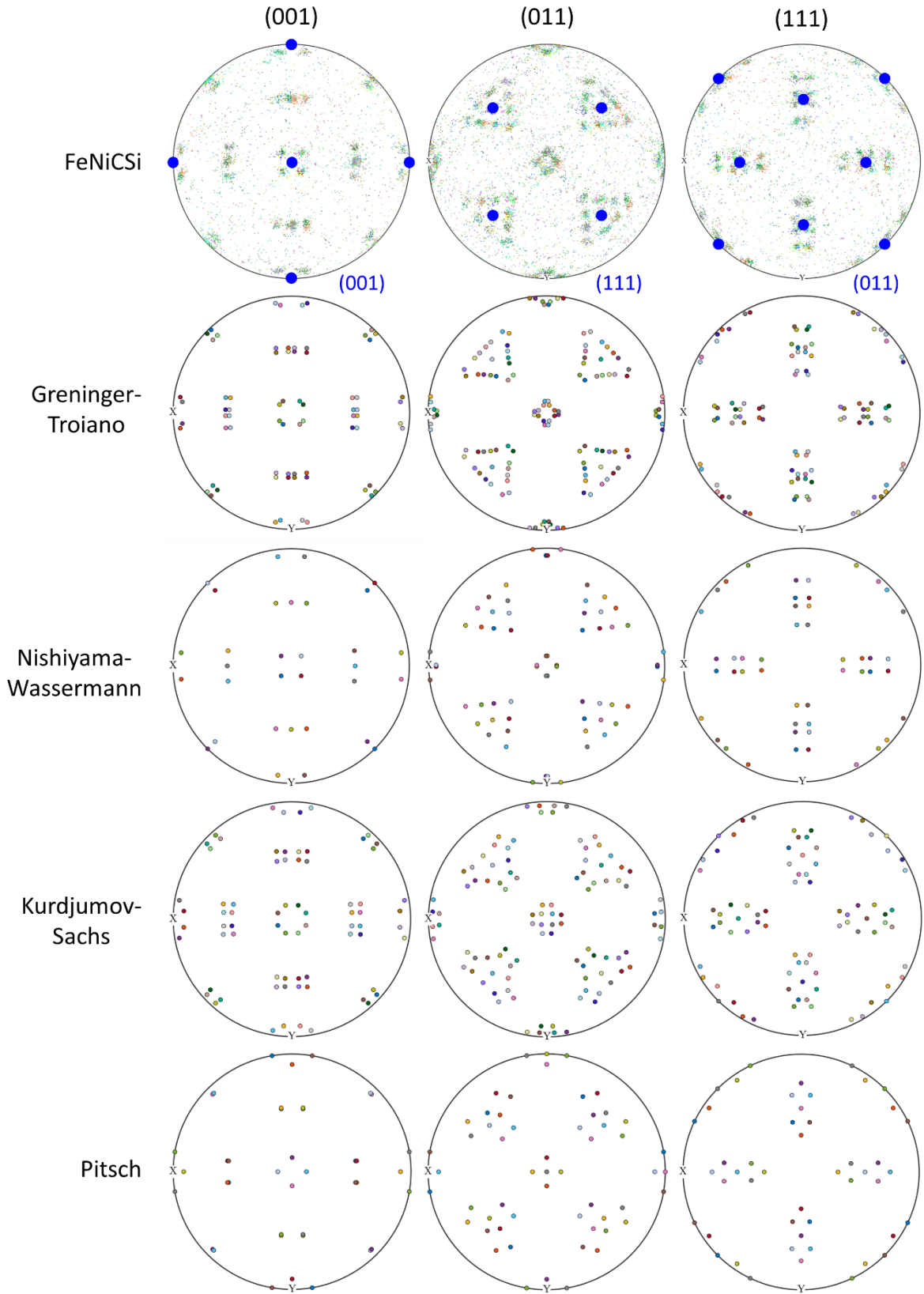

*Fig. 5 Pole figures calculated for different theoretical orientation relationships, namely Kurdjumov-Sachs (KS), Nishiyama-Wasserman (NW), Greninger-Troiano (GT), Pitsch and experimentally obtained data exemplary shown for FeNiCSi showing the reference austenite orientation in blue.*

An example for the semi-automatic selection of grains is depicted in Fig. 6. The blue orientations in the pole figures of FeNiCSi correspond to the austenite grains after rotation into the reference position. Likewise, the rotation angle between the martensite and austenite orientations are utilised to identify the variants. The lowest rotation angle deviation and thus the variant is determined by comparison with martensite variants previously calculated based on the closest matching OR (Table 2), as exemplified in Fig. 6d).

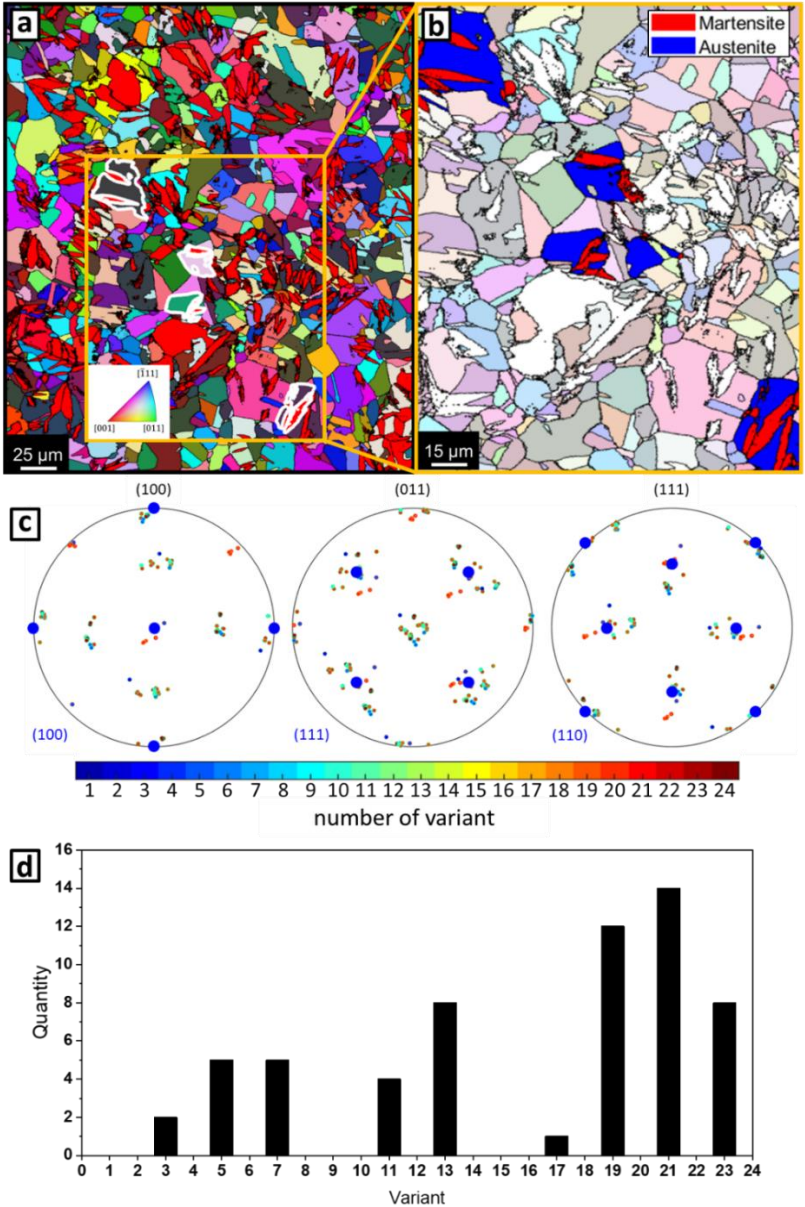

*Fig. 6 a) Selection of grains to be analysed (white) with respect to their orientation relationship or variants. b) Extraction and highlighting of the selected grains visualised in c) for martensite with respect to the rotated austenite orientations shown in blue in the pole figure. d) Histogram of the variants based on the smallest deviation of the rotation angle of the exemplary martensite variants from the extracted grains with respect to the possible simulated variants (here GT).*

*Table 2 Calculated variants from the GT OR*

| Variant No. | Mis. Angle from V1 | Rotation axis from V1 | φ₁ | Φ | φ₂ |
|---|---|---|---|---|---|
| 1 | 0 | - | 324.79 | 170.10 | 191.32 |
| 2 | 55.26 | [-0.71 0.00 0.71] | 98.12 | 84.31 | 136.13 |
| 3 | 60 | [-0.71 0.00 0.71] | 351.88 | 95.70 | 316.13 |
| 4 | 4.74 | [-0.71 0.00 0.71] | 125.21 | 9.90 | 11.32 |
| 5 | 60 | [-0.71 0.00 0.71] | 275.75 | 98.08 | 46.94 |
| 6 | 60.25 | [-0.53 0.53 0.66] | 174.25 | 81.92 | 226.94 |
| 7 | 50.67 | [-0.64 0.42 0.64] | 261.88 | 95.70 | 316.13 |
| 8 | 16.61 | [-0.69 0.23 0.69] | 35.21 | 9.90 | 11.32 |
| 9 | 55.24 | [-0.67 0.23 0.71] | 185.75 | 98.08 | 46.94 |
| 10 | 50.14 | [-0.47 0.57 0.68] | 84.25 | 81.92 | 226.94 |
| 11 | 13.99 | [-0.55 0.06 0.83] | 234.79 | 170.10 | 191.32 |
| 12 | 50.65 | [-0.67 0.26 0.70] | 8.12 | 84.31 | 136.13 |
| 13 | 13.99 | [-0.06 0.55 0.83] | 54.79 | 170.10 | 191.32 |
| 14 | 50.14 | [-0.57 0.47 0.68] | 188.11 | 84.31 | 136.13 |
| 15 | 52.21 | [-0.21 0.65 0.73] | 81.88 | 95.70 | 316.13 |
| 16 | 11.60 | [-0.69 0.19 0.69] | 215.21 | 9.90 | 11.32 |
| 17 | 49.64 | [-0.60 0.52 0.60] | 5.75 | 98.08 | 46.94 |
| 18 | 56.85 | [-0.19 0.66 0.72] | 264.25 | 81.92 | 226.94 |
| 19 | 55.24 | [-0.23 0.67 0.71] | 354.25 | 81.92 | 226.94 |
| 20 | 50.65 | [-0.26 0.67 0.70] | 95.75 | 98.08 | 46.94 |
| 21 | 19.71 | [-0.14 0.00 0.99] | 305.21 | 9.90 | 11.32 |
| 22 | 56.84 | [-0.66 0.19 0.72] | 171.88 | 95.69 | 316.13 |
| 23 | 52.21 | [-0.65 0.21 0.73] | 278.11 | 84.31 | 136.13 |
| 24 | 19.80 | [-0.20 0.00 0.98] | 144.79 | 170.10 | 191.32 |

## 4 Discussion

An approach to compare the similarity of experimental data with potential orientation relationships is the analysis of the misorientation angles. The distributions of the misorientation angles of the martensite/austenite-phases are depicted in Fig. 7 within a range of ± 5 ° for FeNiC and FeNiCSi, respectively. The martensite-austenite misorientation profiles show clear peaks at a misorientation angle of about 44.05 °, indicating that the Greninger-Troiano orientation relationship is matching best to the experimental data.

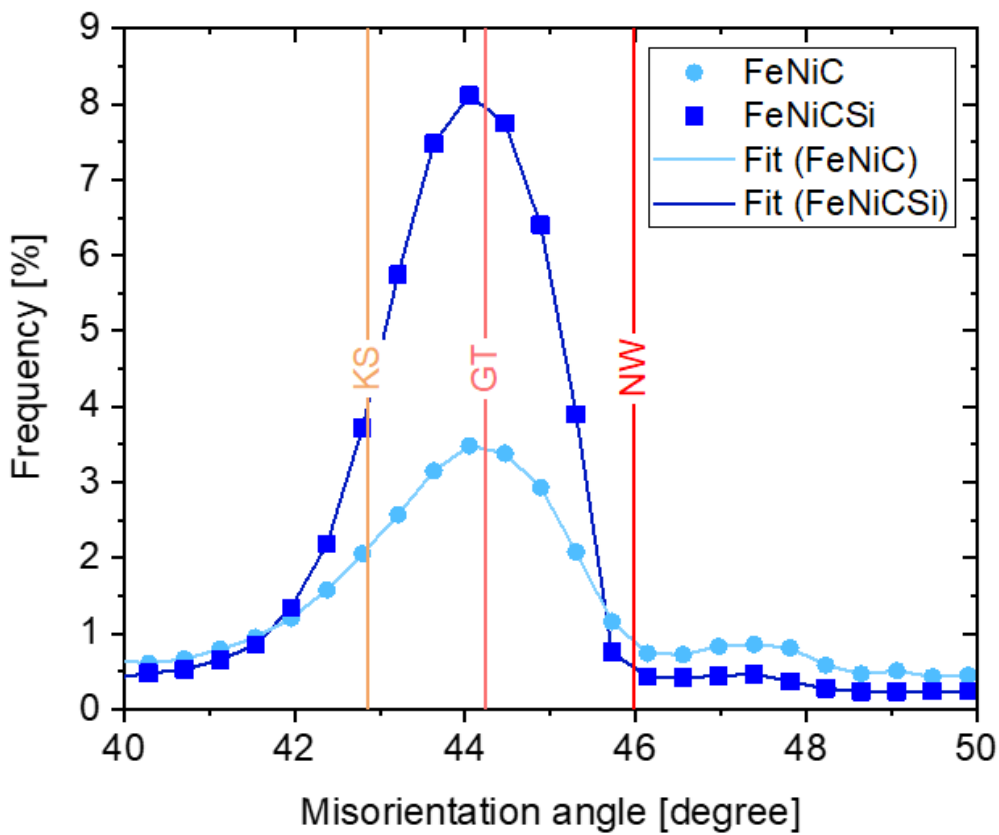

*Fig. 7 Misorientation angle distribution for FeNiC and FeNiCSi.*

Specifically, the mean misorientation angle of FeNiC and FeNiCSi amounts to 44.43°, very close to the misorientation postulated by Greniger and Troiano, which lies between the KS and the NW orientation relationships (Table 3, Fig. 7) [35].

*Table 3 Overview of the different ORs between fcc and bcc crystals and the respective misorientation axis/angle.*

| OR | <uvw> | $\omega_{min}$ |
|---|---|---|
| Bain | <100> | 45° |
| NW | <0.98 0.08 0.20> | 45.98° |
| KS | <0.97 0.18 0.18> | 42.85° |
| GT | <0.97 0.19 0.13> | 44.23° |
| Pitsch | <0.08 0.20 0.98> | 45.98° |

The analysis of the misorientation angle is not always sufficient to accurately determine an OR. To confirm this finding, a subsequent pole figure analysis can be performed. As an approach to validate the obtained OR, normalised cross correlation (NCC), an inverse Fourier transform of the convolution of the Fourier transform of, e.g., two images, can be applied [36, 37]. The normalisation is then done using the local cumulative sums and standard deviations. This type of correlation helps to ascertain the agreement between two data sets and has been applied to compare the simulated and experimental pole figures of the automatic selection mode visually in order to obtain quantitative correlation coefficients. In Fig. 8a)-c) the subtracted simulated and experimental (001), (011) and (111) pole figures of FeNiCSi are depicted. In addition, the correlation factors of the possible ORs given in Table 1 are shown in Fig. 8d), indicating that the dominant OR in both alloys corresponds to the GT-OR with the highest correlation factors. The effects of Si on the initial austenite grain sizes and hence on the thermally transformed martensites have no major impact on the determination of the OR as GT for both, FeNiC and FeNiCSi, in this work.

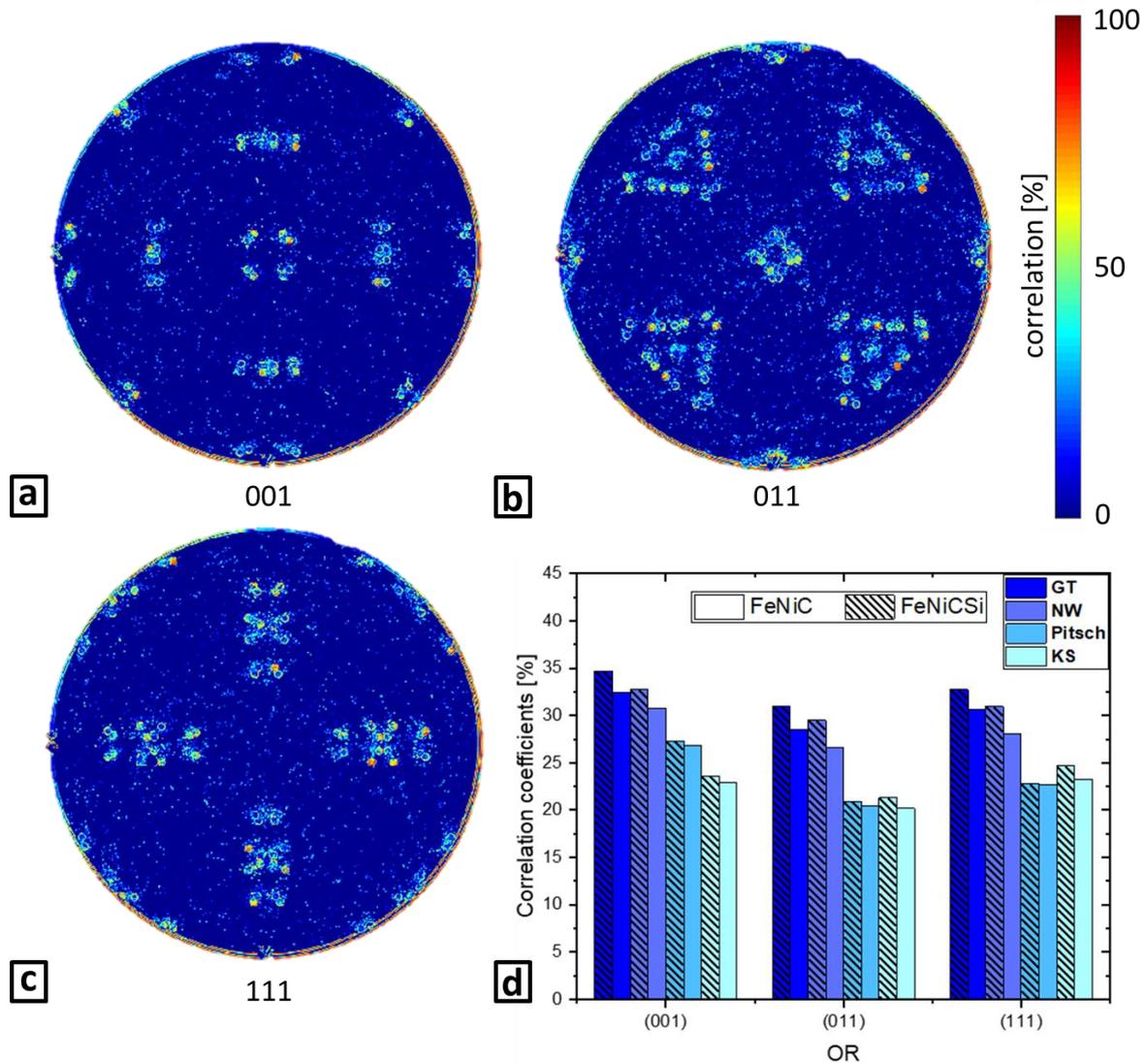

*Fig. 8 Subtraction of simulated (here GT) from experimental a) (001) b) (011) and c) (111) pole figures for FeNiCSi; d) Correlation coefficients of FeNiC and FeNiCSi, respectively, with the simulated pole figures to enable quantitative analysis.*

Another approach to statistically analyse the experimentally measured with the simulated pole figures, the orientation data were compared in terms of the minimum deviation of the rotation angles between the measured orientation and all martensite variants of the respective ORs according to a certain threshold value. Fig. 9 a) displays the experimental raw data of the (011) pole figure, while b) and c) present the filtered experimental data for a minimum rotation angle deviation with a threshold of <10 ° and <5 °, respectively. For a threshold value of <5°, about ~86 % of all orientations of the original EBSD map are preserved, allowing a statistical analysis of the large majority of the experimental data. As a result of multiple phases or grain boundaries in material systems, the EBSD measurement is partially unable to assign the correct diffraction patterns unambiguously. In comparison of all ORs at the different threshold values and the average values of both alloys, the smallest rotational deviations are obtained for the GT OR, demonstrating that the misorientation distribution and combined pole figure comparison are suitable to determine orientation relationships.

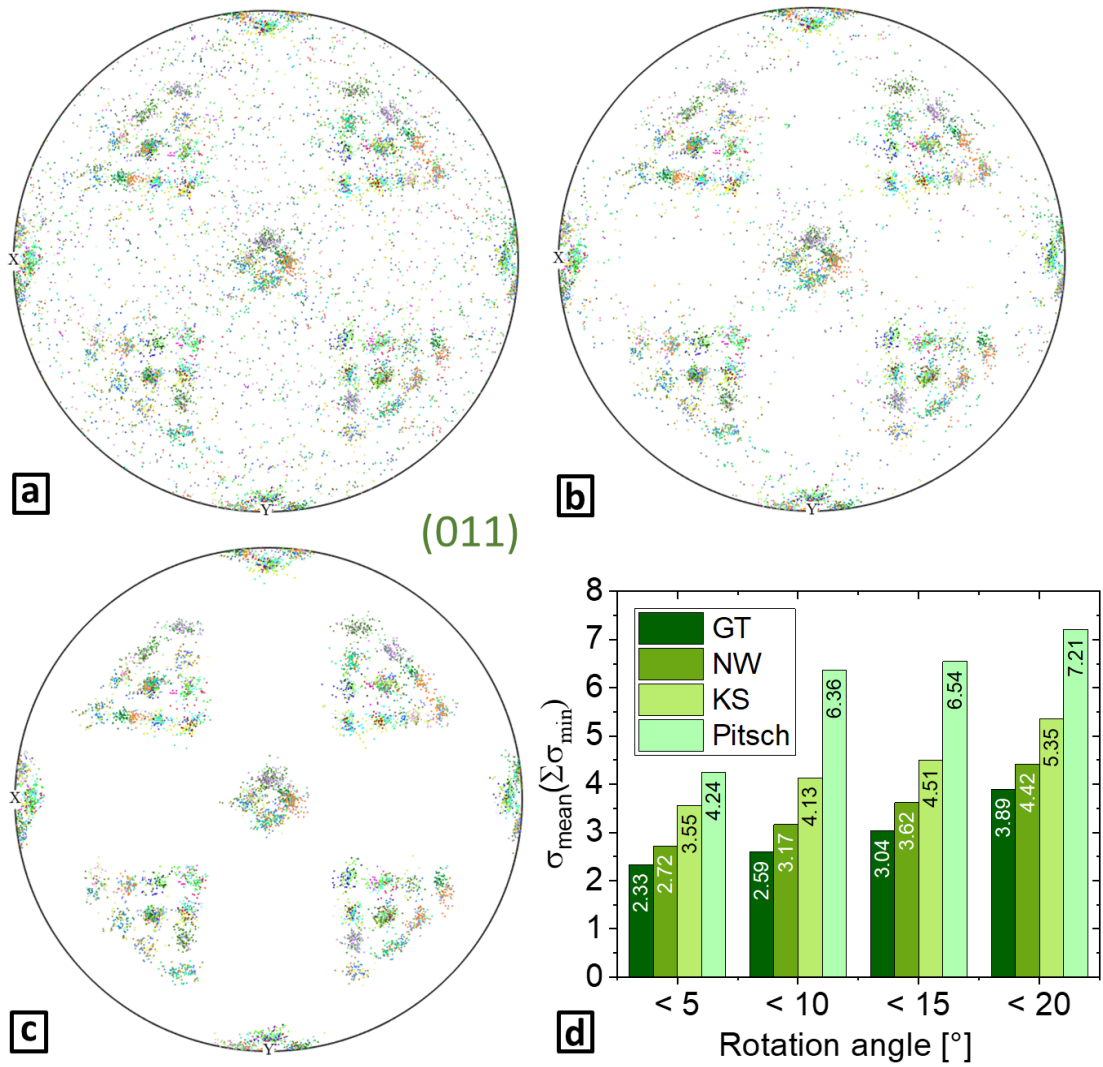

*Fig. 9 Pole figures of the raw and filtered orientation data measured for FeNiCSi and the statistical evaluation of the orientation data exemplary for the (011) pole figure. a) raw pole figure, b) pole figure filtered by rotation angle deviation <10° and c) pole figure filtered by rotation angle deviation <5° and d) histograms showing the mean values for each set of minimum rotation angle deviation for each OR revealing that the lowest rotational deviations are present for the GT orientation relationship.*

Table 4 presents the correlation coefficients and rotational deviations of FeNiC and FeNiCSi when considering the possible orientation relationships given in Table 1. The higher the correlation coefficient of the experimental data with one of the possible ORs, the higher the similarity between them. Similarly, a lower rotational deviation indicates higher similarity. In the present examples, all methods to determine the OR show the highest match with the GT-OR.

*Table 4 Mean correlation coefficients and mean orientation angle deviation calculated for FeNiC and FeNiCSi.*

| OR | NW | KS | GT | Pitsch |
|---|---|---|---|---|
| Mean $C_{corr}$ (FeNiC) | 28.50 | 24.05 | 30.51 | 23.33 |
| Mean $C_{corr}$ (FeNiCSi) | 31.05 | 25.41 | 32.84 | 23.66 |
| Mean σ (FeNiC) | 3.48 | 4.39 | 2.97 | 6.09 |
| Mean σ (FeNiCSi) | 3.70 | 4.54 | 3.12 | 6.04 |

The experimental data of the martensite variants of both, FeNiC and FeNiCSi, presented in austenite reference pole figures in this work (Fig. 5) provide a statistically close correspondence with the GT OR in comparison with other established ORs like KS, NW or Pitsch. It has been shown that the code developed for identifying orientation relationships gives quite unambiguous results for reasonably accurate EBSD and / or TKD datasets. The main component is the automatically generated statistical evaluation of all grain orientations within an entire EBSD / TKD measurement and relating them to a predefined orientation relationship. In addition, the reference systems can be modified. In principle, the code was designed as a workflow for several use cases with some defined pre-settings, but it is also intended to be kept as user-friendly as possible, which means that functional components can be adjusted in the respective source codes. It is mainly optimised for two-phase systems, but can also visualise at least individual orientation relationships in multi-phase systems through the manual selection of grains. If the EBSD / TKD map quality is insufficient for a statistical evaluation, manual grain selection may be an option to determine the orientation relationship of individual areas. Since the local misorientations within individual grains can diverge significantly, the manual mode enables the visualisation of the orientation relationship of single, acquired orientation spots or areas within a grain. Although this method could also be useful for e.g. phase shape memory alloys (Ti-Ni [38, 39]) or intermetallic phases (Mg-Ag-Al [40]), it has so far only been investigated for the material systems described in this work.

The aforementioned concept is limited to use cases in which grains of both phases and their boundaries remain for OR analysis. Thus, it is not applicable after a complete phase transformation. In an Fe-24Ni-0.3C alloy, by comparing ultrafine with coarse grained austenitic microstructure, it was observed that the martensite/austenite interface exhibited the GT OR on both sides in coarse grains and in addition the K- S OR on the outer side of the martensite in ultrafine grains [41]. Micromechanical effects, such as strain fields or distortions between neighbouring martensite variants, could have an influence on the crystal orientation of the martensite variants, which complicates a precise differentiation of individual ORs [42]. In an Fe-33Ni alloy, it was found that the OR in lenticular martensite varied from the midrib (GT) to the austenite/martensite interface (KS) [15]. However, in an Fe-31%Ni-0.01C alloy, where partially transformed lenticular martensite in conjunction with austenite was present, it was reported that although there is a scatter of orientations, near both, the interface and the midrib, the misorientations are closer to the GT orientation, resulting in the OR being the same between the midrib and the interfacial region [43]. Due to the potential existence of several ORs in one material and to enable a more representative demonstration of the ORs, further work is required.

All the codes discussed in this work are maintained in an online repository available at the following reference: [34].

# 5 Conclusions

A Matlab® code based on the MTEX toolbar for the quantitative identification of orientation relationships with statistical relevance has been developed. As an example, the OR between retained austenite and martensite was determined and statistically evaluated. From our study the following conclusions have been derived:

- The code contains different modes that can be used for an in-depth analysis of obtained ORs, specifically, automatic, semi-automatic and manual selection of grains or microstructure areas.
- The automatic mode allows a statistically relevant and fully automatic evaluation of the orientation relationship of a large range of grains and thereby allows the determination of the predominant OR. This is done by comparing the pole figures of the corresponding phases of interest, e.g., retained parent austenite grains and the adjacent martensite grains, with a calculated theoretical orientation relationship.
- Additionally, the manual evaluation mode provides local insights into the precise OR for chosen grains or individual interfaces, in the present case a determination of the occurring martensite variants was possible. The selection of individual measurement points instead of whole grains is also possible and could be used to evaluate the behavioural impact of intragranular misorientations on the OR or can be applied if the indexing rate of the EBSD data is low.
- Exemplarily, the code was applied to an austenitic-martensitic steel where the successful identification and evaluation of the OR between austenite and martensite was demonstrated. The comparison of the misorientation angle distribution with the misorientation angles of specific ORs supports the identification of the Greninger-Troiano OR for both FeNiC and FeNiCSi. This was further confirmed by an image correlation algorithm, comparing calculated theoretical variants with the experimental pole figures.
- For quantitative analysis of the observed OR, the minimum rotation angle deviations between the experimental and the theoretically calculated orientation relationships were determined for different ORs. This method was successfully applied to confirm the previously identified Greninger-Troiano OR for the present example case.
- Although the capabilities of the developed code on the example of martensite and austenite in steels have been demonstrated, it can be used for many other applications where an orientation relationship between different phases is relevant. An adaptation of the present method for the investigation of ORs in other material systems such as intermetallic phases, shape memory alloys, thin films or other composite materials could provide insights into the existing ORs.

## Acknowledgement

The authors gratefully acknowledge funding by the Deutsche Forschungsgemeinschaft (DFG, German Research Foundation) through project 406912286 (C-Tram).

## Data Availability

The raw/processed data required to reproduce these findings cannot be shared at this time as the data also forms part of an ongoing study.